\documentclass[]{aa}  

\usepackage{graphicx}
\usepackage{subcaption}
\usepackage{caption}
\usepackage{float}
\usepackage{lipsum}

\usepackage{txfonts}
\usepackage[colorlinks=true, allcolors=blue]{hyperref}
\usepackage{comment}
\begin{document}

   \title{\textsc{The Three Hundred} project: Estimating the dependence of gas filaments on the mass of galaxy clusters}

   \author{Sara Santoni
          \inst{\ref{sap},\ref{uamdip},\ref{eduam}}
          \and Marco De Petris
          \inst{\ref{sap}}
          \and Gustavo Yepes
          \inst{\ref{uamdip},\ref{uamciaff}}
          \and Antonio Ferragamo
          \inst{\ref{sap},\ref{nap}}
          \and Matteo Bianconi
          \inst{\ref{nott}}
          \and Meghan E. Gray
          \inst{\ref{nott}}
          \and Ulrike Kuchner
          \inst{\ref{nott}}
          \and Frazer R. Pearce
          \inst{\ref{nott}}
          \and Weiguang Cui
          \inst{\ref{uamdip},\ref{uamciaff},\ref{edin}}
          \and Stefano Ettori
          \inst{\ref{inafbo},\ref{infnbo}}
          }

   \institute{Dipartimento di Fisica, Sapienza Università di Roma, Piazzale Aldo Moro 5, I-00185 Rome, Italy\\
    \email{sara.santoni@uniroma1.it} \label{sap}
        \and {Departamento de Física Teórica, Facultad de Ciencias, Universidad Autónoma de Madrid, Modulo 8, E-28049 Madrid, Spain} \label{uamdip}
        \and{Escuela de Doctorado UAM, Centro de Estudios de Posgrado, Universidad Autónoma de Madrid, E-28049, Madrid, Spain} \label{eduam}
        \and {Centro de Investigación Avanzada en Física Fundamental (CIAFF), Facultad de Ciencias, Universidad Autónoma de Madrid, E-28049, Madrid, Spain}\label{uamciaff}
        \and{Dipartimento di Fisica Ettore Pancini, Università degli Studi di Napoli Federico II, Via Cintia 21, I-80126 Napoli, Italy} \label{nap}
        \and {School of Physics and Astronomy, University of Nottingham, Nottingham NG7 2RD, UK} \label{nott}
        \and {Institute for Astronomy, University of Edinburgh, Edinburgh EH9 3HJ, UK} \label{edin}
        \and{INAF, Osservatorio di Astrofisica e Scienza dello Spazio, via Piero Gobetti 93/3, I-40129 Bologna, Italy} \label{inafbo}
        \and{INFN, Sezione di Bologna, viale Berti Pichat 6/2, I-40127 Bologna, Italy} \label{infnbo}
}

   \date{Received ...; accepted ...}

 
  \abstract
   {Galaxy clusters are located in the densest areas of the universe and are intricately connected to larger structures through the filamentary network of the cosmic web. In this scenario, matter flows from areas of lower density to higher density. As a result, the properties of galaxy clusters are deeply influenced by the filaments that are attached to them, which are quantified by a parameter known as connectivity.}
   {We explore the dependence of gas-traced filaments connected to galaxy clusters on the mass and dynamical state of the cluster. Moreover, we evaluate the effectiveness of the cosmic web extraction procedure from the gas density maps of simulated cluster regions.}
   {Using the DisPerSE cosmic web finder, we identify filamentary structures from the 3D gas particle distribution in 324 simulated regions of $30 \, h^{-1}$ Mpc side from \textsc{The Three Hundred} hydrodynamical simulation at redshifts $z=0$, 1, and 2. We estimate the connectivity at various apertures for $\sim3000$ groups and clusters spanning a mass range from $10^{13} \, h^{-1} \, M_{\odot}$ to $10^{15} \, h^{-1} \, M_{\odot}$. Relationships between connectivity and cluster properties like radius, mass, dynamical state, and hydrostatic mass bias are explored.}
   {We show that the connectivity is strongly correlated with the mass of galaxy clusters, with more massive clusters being on average more connected. This finding aligns with previous studies in the literature, both from observational and simulated datasets. Additionally, we observe a dependence of the connectivity on the aperture at which it is estimated. We find that connectivity decreases with cosmic time, while no dependencies on the dynamical state and hydrostatic mass bias of the cluster are found. Lastly, we observe a significant agreement between the connectivity measured from gas-traced and mock-galaxy-traced filaments in the simulation.}
   {}

   \keywords{large-scale structure of Universe --
                Galaxies: clusters: general --
                Methods: numerical -- 
                Methods: statistical
               }

   \maketitle
%

\section{Introduction}
Clusters of galaxies, considered the largest gravitationally bound structures in the universe, are located at the nodes of an intricate web-like structure known as the ‘cosmic web’ \citep{Bond:1996}. This cosmic pattern of dark matter (DM) and baryons and the formation of large-scale structures are predicted as a consequence of the non-linear growth of primordial density perturbations \citep{Zel'dovich:1970,Peebles:1980}. In this framework, matter flows from underdense regions to overdense structures, resulting in compact galaxy clusters at the nodes of elongated filamentary structures and filaments at the intersection of sheet-like formations, known as walls, surrounded by vast void regions. 

The cosmic web has been observed by several galaxy surveys, such as the Sloan Digital Sky Survey \citep[SDSS;][]{York:2000} and the 2dF Galaxy Redshift Survey \citep[2dFGRS;][]{Colless:2001}. Furthermore, the properties of the cosmic web and its components have been identified through several large-scale numerical simulations, both N-body and hydrodynamical, such as the Millennium simulation \citep{Springel:2005}, the MilleniumTNG project \citep{Hernandez-Aguayo:2023}, the Illustris project \citep{Vogelsberger:2014}, IllustrisTNG \citep{Nelson:2019}, the EAGLE project \citep{Schaye:2015}, and the SIMBA project \citep{Dave:2019}.

Depending on the scientific goal of the study, different features of the cosmic web can be examined, often without relying on unequivocal definitions. Consequently, several algorithms have been developed to identify the cosmic web, some tailored for simulations and others for observational data, as is illustrated in the exhaustive review by \citet{Libeskind:2018}. The different algorithms generally converge on the identification and characterisation of specific features, but they reflect the complexity involved in the study and definitions of these structures. 

In particular, the most used extraction methods can be grouped as graph and percolation techniques (e.g. MST, \citet{Alpaslan:2014} and T-Rex, \citet{Bonnaire:2020}), stochastic methods (e.g. Bisous model, \citet{Tempel:2016}), methods based on the Hessian of the density, tidal, or velocity fields (e.g. T-web, \citet{Forero-Romero:2009}, V-web, \citet{Hoffman:2012}), and, among these, those that follow a multi-scale approach (e.g. MMF, \citet{Aragon-Calvo:2007}, NEXUS and NEXUS+, \citet{Cautun:2013}. Moreover, the cosmic web can be identified with topological methods (e.g. Spineweb, \citet{Aragon-Calvo:2010a} and DisPerSE, \citet{Sousbie:2011a,Sousbie:2011b}) and, lastly, phase-space methods (e.g. ORIGAMI, \citet{Falck:2012}. 

In addition to the galaxy clusters, filaments constitute the densest regions of the cosmic web, containing approximately $50 \% $ of the matter in the universe \citep{Cautun:2014}.
Filaments serve an important function in the accretion phase of galaxy clusters, acting as pathways for matter transportation towards the centre of the node to which they are connected.
The connectivity of the cosmic web, which measures the number of filaments connected to a cluster, offers a method of quantifying the interaction between clusters and the filaments surrounding them, enabling the assessment of structural evolution \citep{Codis:2018}. Several works in the literature have been investigating the correlation between connectivity and a cluster's properties, such as its mass and its dynamical state \citep[see e.g.][]{DarraghFord:2019,Sarron:2019,Malavasi:2020b,Gouin:2021,Galarraga-Espinosa:2024}. 

Furthermore, several studies proved that filaments play a significant role in the so-called ‘pre-processing’ of the galaxies that reside within the filamentary structures. Filaments affect galaxy evolution and galaxy properties, before entering the cluster, through several processes as ram-pressure, tidal effects, and mergers (see the reviews of \citealt{Boselli:2006,Boselli:2014} and more recent works e.g. \citealt{Kraljic:2018, Winkel:2021, Hellwing:2021, Song:2021,Donnan:2022, Kotecha:2022,Kuchner:2022,Malavasi:2022,Galarraga-Espinosa:2024, Bulichi:2024}). Therefore, the outskirts of clusters provide a distinctive environment for examining matter accretion dynamics and galaxy properties. 

Unlike the galaxy content of filaments, which has been extensively identified with galaxy surveys, the detection of the gas component has proven more challenging. Hydrodynamical simulations \citep[see e.g.][]{Bond:1996,Dave:2001,Shull:2012,Tuominen:2021} predict that the filamentary structures of the cosmic web host a large fraction of the missing baryons\footnote{When compared to the concordance cosmology model, we observe a $\approx30 \%$ deficit in the census of observed baryons from low-redshift sources, such as galaxies, groups, and clusters, in the intergalactic and circumgalactic medium.} in a low-density and warm-hot state known as the warm-hot intergalactic medium (WHIM), with a density of $\delta_b \approx 10-100$ and a temperature in the range of $10^5 - 10^7 $ K. 

The hot phase of WHIM, with $\log T(K) \approx 5.5-7$, is not observable through absorption, being fully ionised, contrary to the low-temperature phase, with $\log T(K) \approx 5-5.5$, which has been detected several times through metal absorption lines \citep{Danforth:2016}. Despite having a faint emission, several attempts have been made to detect the warmest phase of WHIM through direct X-ray observations \citep[see e.g.][]{Eckert:2015,Nicastro:2018} and the Sunyaev-Zel'dovich effect, employing stacking methods \citep{Tanimura:2019,Tanimura:2020,Tanimura:2022} or looking at the densest regions between two close galaxy clusters \citep{Bonjean:2018,Hincks:2022}. 

In this work, we use the outputs of \textsc{The Three Hundred} \citep{Cui:2018b} hydrodynamical simulations to extract the cosmic web in the surrounding volumes of galaxy clusters, with a specific focus on the filamentary structures, from the gas particle distribution using the cosmic web finder DisPerSE. By analysing the connectivity estimates of simulated galaxy clusters, this investigation explores the influence of filaments on the main properties of galaxy clusters, such as their mass and dynamical state, and assesses the cosmic web extraction from diffuse gas particles within the simulated \textsc{The Three Hundred} regions. 

This paper is structured as follows. Sect. \ref{sec:2} provides a detailed overview of \textsc{The Three Hundred} simulation and the theoretical framework behind the cosmic web finder DisPerSE. The process of creating the gas density maps and calibrating DisPerSE is discussed in Sect. \ref{sec:3}. In Sect. \ref{sec:4}, we present the results of the connectivity analysis, while in Sect. \ref{sec:5} we investigate the effect of cosmic web tracers by comparing the gas-traced filaments to mock-galaxy-traced filaments. Finally, Sect. \ref{sec:6} summarises the main results of our study. 

\section{Simulation and methods}
\label{sec:2}
\subsection{\textrm{\textsc{The Three Hundred}} project}
The dataset analysed in this work was extracted from \textsc{The Three Hundred} project\footnote{\href{https://www.nottingham.ac.uk/~ppzfrp/The300/}{https://the300-project.org}} \citep{Cui:2018b}. \textsc{The Three Hundred} hydrodynamical simulation consists of 324 zoom-in re-simulated regions centred on the most massive haloes selected at $z=0$ from the MultiDark Planck 2 (hereafter as MDPL2) simulation \citep{Klypin:2016}. MDPL2 is a DM-only cosmological simulation, consisting of a periodic cube with a side length of $1 \, h^{-1}$ Gpc with $3840^3$ DM particles, each with a mass of $1.5 \times 10^9 \, h^{-1} \, M_{\sun}$. The cosmological parameters used in MDPL2 and \textsc{The Three Hundred} are those reported in the 2015 \textit{Planck} data release \citep{PlanckCollaboration:2016} ($h = 0.678$, $n_s = 0.96$, $\sigma_8 = 0.823$, $\Omega_{\Lambda} = 0.693$, $\Omega_{\textrm{M}} = 0.307$, and $\Omega_{\textrm{B}} = 0.048$). 

The 324 spherical regions of \textsc{The Three Hundred}, centred on the most massive haloes ($M_{vir} > 8 \times 10^{14} \, h^{-1} \, M_{\sun}$), have a radius of $15 \, h^{-1}$ Mpc. The DM particles within these regions are split into separate DM and gas particles, according to the cosmological baryonic mass fraction, with masses of $m_{\textrm{DM}} = 12.7 \times 10^8 \, h^{-1} \, M_{\sun}$ and $m_{\textrm{gas}} = 2.36 \times 10^8 \, h^{-1} \, M_{\sun}$. 

In this analysis, we have used the regions re-simulated with the smoothed particle hydrodynamics (SPH) code \textsc{Gadget-X} \citep{Springel:2005,Beck:2016} -- a modified version of the \textsc{Gadget3} code -- which implements radiative cooling and star formation \citep{Springel:2003}, chemical stellar evolution \citep{Tornatore:2007}, supermassive black hole (BH) accretion, and active galactic nuclei and supernovae feedback \citep{Steinborn:2015}. The haloes and subhaloes within each region were identified with the publicly available\footnote{\href{http://popia.ft.uam.es/AHF/}{http://popia.ft.uam.es/AHF/}} Amiga Halo Finder  \citep[AHF;][]{Knollmann:2011}, which includes DM, gas, stars, and BHs in the density peak finding process and estimates the co-ordinates and properties of the clusters, such as their mass, radius, and luminosities at various overdensities. The simulation runs are stored in 129 snapshots, from redshift $z=17$ to $z=0$. In this work, we analyse the galaxy clusters' catalogues at redshifts $z=0$, $z=1.03$, and $z=2.02$, which span, at $z=0$, a mass range of $10^{13} \leq M_{200} \, h^{-1} \, M_{\sun} \leq 2.62 \times 10^{15}$.

\subsection{DisPerSE}
The cosmic web in the simulated regions was extracted using the publicly available\footnote{\href{https://www2.iap.fr/users/sousbie/web/html/indexd41d.html}{https://www2.iap.fr/users/sousbie/web/html/indexd41d.html}} Discrete Persistent Structure Extractor \citep[DisPerSE;][]{Sousbie:2011a,Sousbie:2011b}. DisPerSE is a topological structure finder, designed to identify multi-scale structures. It relies on discrete morse theory \citep{Forman:1998, Gyulassy:2008} to identify the critical points of density fields. 
The noise introduced into the detection by the finite sampling of the density field was reduced with the approach of persistent homology theory \citep{Edelsbrunner:2002} and topological simplification \citep{Edelsbrunner:2002,Gyulassy:2008}. 

In short, from a given 2D or 3D distribution of tracers, DisPerSE provides the position of the extreme points -- where the gradient of the field is null -- and the integral lines that connect them, which form the filaments. The starting point of DisPerSE can either be a catalogue of the tracers' co-ordinates (e.g. galaxies or haloes) or a pixelated density map, as is used in this work. In the first case, the density field is extracted from the Delaunay tessellation of the distribution with the Delaunay tessellation field estimator \citep[DTFE;][]{Schaap:2000,vandeWeygaert:2009}.

From the density field, the identified extreme points were classified as maxima, minima, saddle, and bifurcation points; that is, where a filament splits in two. The ridges were presented as a set of segments that connect two extreme points -- usually a maximum and a saddle point -- following the density field. The robustness of filaments -- that is, their reliability -- was quantified through a threshold parameter called ‘persistence cut’, defined as the absolute difference in the density value of the two critical points at the filament's extremes. The choice of the persistence allows for low-significance and noisy structures to be filtered out.

\section{Cosmic web extraction}
\label{sec:3}
\subsection{DisPerSE calibration}

In order to extract a significant filamentary skeleton from the density distribution, a combination of parameters -- the Gaussian smoothing of the input density map and the persistence cut of DisPerSE -- needs to be tuned. The first step of the cosmic web extraction is the smoothed gas density field reconstructed from the regions of the hydrodynamical simulation. \textsc{The Three Hundred} gas-traced cosmic web has been widely analysed in previous works; namely, to study the properties of filaments \citep{Rost:2021,Rost:2024} and as a benchmark for the identification of galaxy-traced filaments \citep{Kuchner:2020,Kuchner:2021}.
In this work, we have chosen to analyse the filamentary structures traced from the gas particle distribution because we plan, in future studies, to use 3D gas filaments and their projection as a benchmark for the detection of filaments from mock Compton-\textit{y} maps generated in \textsc{The Three Hundred} simulation, as is described in \citet{Cui:2018b}. 

\begin{figure}
   \centering
   \includegraphics[width=8.8cm]{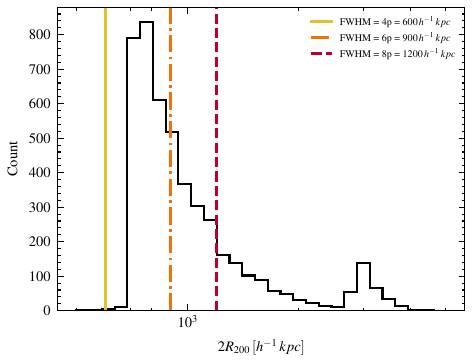}
   \caption{Diameter $2 R_{200}$ distribution of galaxy clusters at $z=0$. The vertical lines represent the smoothing kernel FWHM at 4, 6, and 8 pixels, respectively, in yellow (solid), orange (dash-dotted), and red (dashed).}
              \label{Fig:dist-radius-fwhm}%
\end{figure}

The gas particles of each simulated region were first binned in a 3D grid of $30 \, h^{-1}$ Mpc per side, with a pixel resolution of $150 \, h^{-1}$ kpc. 
Then, to avoid sharp variations from one pixel to another, we applied a Gaussian kernel to smooth the density field. As the smoothing kernel of the map increases, only the more distinct structures will be identified by the filament finder, leading to a network that only features robust critical points and filaments. For this analysis, we applied a Gaussian kernel with a full width at half maximum (FWHM) of 4 pixels, equivalent to $600 \, h^{-1}$ kpc. We opted for this specific kernel size, ensuring it remains smaller than the diameter of the galaxy clusters that we analysed, to preserve the resolution of these structures, as is shown in Fig. \ref{Fig:dist-radius-fwhm}, where we compare the chosen value and two larger smoothing kernels to the distribution of the diameters of clusters at $z=0$. 

In addition to the smoothing level of the density map, the robustness of the extracted network depends also on the DisPerSE parameter ‘persistence’. The effect of the persistence cut on the network is similar to the Gaussian smoothing, with higher values of persistence leading to less critical point pairs, and therefore to the removal of less significant filaments. The optimal value of persistence cut for our dataset is 0.5, which identifies the threshold between significant topological features and noisy ones, as is seen in the DisPerSE Persistence Diagram Viewer (\texttt{pdview}), accessible with option \texttt{-interactive} of the function \texttt{mse}. 
The persistence cut differs from the commonly referenced persistence ratio, expressed as a number of standard deviation $\sigma$ and defined as the ratio of the values of the points in a critical pair. The persistence ratio is often applied to a density field of discrete catalogues (e.g. computed using DTFE). The persistence absolute cut of 0.5 that we apply to 3D gas density maps is comparable to a persistence ratio of $8.5 \sigma$, producing filament skeletons that look visually similar.

The density maps were first analysed by DisPerSE with the function \texttt{mse}, with the options \texttt{-forceLoops}, \texttt{-periodicity 000}, \texttt{-cut 0.5}, and \texttt{-upSkl}, to compute the Morse-smale complexes. During this step, the filaments were extracted and the critical points were identified. Then, the networks were post-treated with the function \texttt{skelconv}, which we implemented with the following options: \texttt{-smooth 1}, \texttt{-breakdown}, \texttt{-rmOutside}, and \texttt{-rmBoundary}. 
Here, the option \texttt{-breakdown} merges the infinitely close pieces of arcs and adds the bifurcations points, while the \texttt{-smooth} option determines the straightness of the filament's trace, while keeping the critical points' co-ordinates fixed. We note that the \texttt{-smooth} function does not affect the extraction and positions of filamentary structures and, although it has a similar name, it is different from the Gaussian smoothing kernel applied to the density maps, which operates on the density field prior to the cosmic web extraction procedure on DisPerSE.

\subsection{Simulated cosmic web network}
\label{sec:3.2}
In the DisPerSE output networks, the galaxy groups and clusters are defined only by the co-ordinates of the maxima (the nodes), while the filaments are defined only by the position of their spines. To associate the extracted skeleton with the density distribution of the simulation, one needs to match the nodes to \textsc{The Three Hundred} galaxy clusters, identified with \textsc{AHF}. While there are several ways to associate the nodes with clusters \citep[for a complete study we refer the reader to][]{Cohn:2022}, in this analysis we required the distance between the co-ordinates of the node and the cluster to be less than the smoothing kernel of $4$ pixels. Where multiple pairs satisfied this condition, we matched the DisPerSE node to the nearest \textsc{AHF} cluster. 

\begin{figure}
   \centering
   \includegraphics[width=8.8cm]{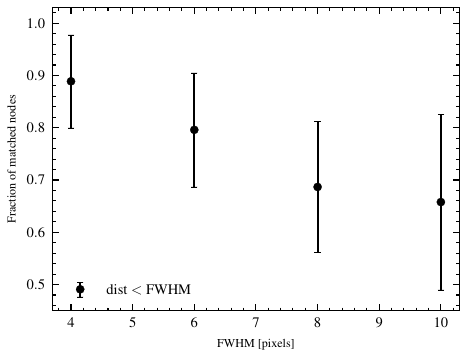}
   \caption{Completeness of the sample. Each dot refers to the completeness of the sample, averaged over the 324 simulated regions, extracted with different smoothing kernels. The error bars represent the standard deviation error.}
              \label{Fig:completeness}%
\end{figure}

To evaluate the accuracy of the DisPerSE catalogue and of the matches, we can compute the completeness of the sample, defined as the number of groups and clusters matched to nodes, normalised by the total number of groups, as in \citet{Cornwell:2024}. The completeness depends on the smoothing scale of the density maps, with a larger and less filtering smoothing kernel resulting in lower completeness of the sample, as is shown in Fig. \ref{Fig:completeness}. Here, we display the completeness of the sample as a function of the Gaussian smoothing FWHM, averaged over the 324 simulated regions. To test only the effect of the smoothing scale on the completeness of the sample, we used the same persistence cut for all the smoothing values, which alters the skeleton in a similar way to the smoothing. This result agrees with \citet{Cohn:2022} and \citet{Galarraga-Espinosa:2024}, where they studied the effect of the smoothing on the completeness of skeletons extracted from DM and galaxy distributions, respectively. 

The sample analysed in this work has an average completeness value of $89 \% $, indicating that almost every group or cluster identified in the simulated regions has a matched DisPerSE node. It is worth noticing that since the completeness is lower than the ideal value of $100 \%$, some clusters remain unmatched by any DisPerSE nodes. This discrepancy might be attributed to their co-ordinates' distance exceeding the threshold we set. This phenomenon could be explained by considering that DisPerSE node co-ordinates are determined solely from the gas-only density distribution, unlike cluster co-ordinates, which also incorporate data from DM particles. Therefore, the different spatial distributions of DM and gas particles near the centre of the regions could affect the peak density finding process \citep{Cui:2018a,Lokken:2023}. 
Conversely, there are DisPerSE nodes without any corresponding clusters. This population of nodes likely represents local maximum points of the density field not classified as haloes by \textsc{AHF} or nodes associated with sub-haloes. 
To avoid possible contamination by low-resolution DM particles near the borders, we considered only the matched nodes inside a sphere of $ 13 \, h^{-1}$ Mpc radius from the centre of each region. Lastly, we note that the galaxy groups and clusters analysed in this work -- except the 324 most massive ones -- reside within the $15 \, h^{-1}$ Mpc environment of very massive galaxy clusters ($M_{vir} > 8 \times 10^{14} \, h^{-1} \, M_{\odot}$).  

\begin{figure}
   \centering
   \includegraphics[width=8.8cm]{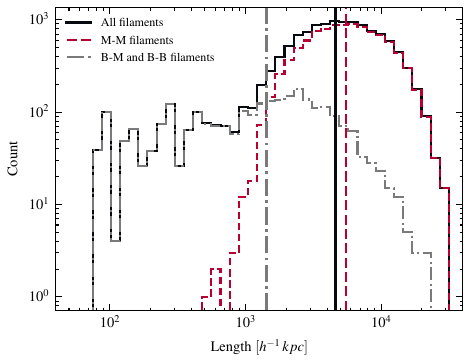}
   \caption{Filaments length distribution. With the solid black line we plot the distribution of length for the whole sample of filaments, with the dashed red line the length distribution of the filaments with two maxima at the extremes, and with the dash-dotted grey line the length of bifurcation-maximum and bifurcation-bifurcation filaments. The vertical lines are the median values of each distribution. }
              \label{Fig:fil-length}%
\end{figure}

The final sample used for this analysis, including all the 324 simulated regions, amounts to  3412 galaxy groups and clusters and $12 \, 643$ filaments at $z=0$. The filament sample includes filaments going from a maximum point to another -- with a saddle point in the middle -- filaments from one maximum point to a bifurcation point and filaments with two bifurcation points at the extremes. The filament length distribution, computed as the sum of the lengths of all segments belonging to each filament, is shown in Fig. \ref{Fig:fil-length}, along with the median values of the distribution. We plot the length distribution of the whole sample in black, the length distribution of maximum-maximum filaments in red, and the length of bifurcation-maximum and bifurcation-bifurcation filaments in grey. 
The mean length of the maximum-maximum filaments is $6.67 \, h^{-1}$ Mpc, comparable to the length of filaments extracted in \citet{Rost:2021} from the same gas particle distribution but with different levels of smoothing of the density map and persistence. Due to the exploration of the limited volume of each region, compared to different studies in the literature, the extracted filaments predominantly belong to the short filament population, as has been shown in previous studies \citep{Bond:2010a, Galarraga-Espinosa:2020}. 

\begin{figure*}[!phtb]
\centering
    \begin{subfigure}[p]{0.75\textwidth}
    \centering
        \includegraphics[width=\linewidth]{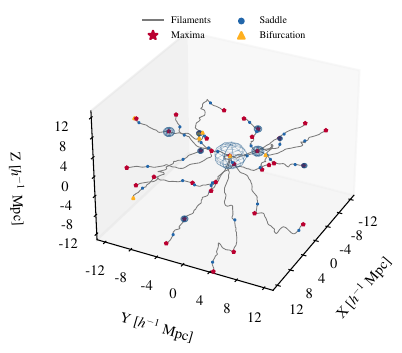} 
        \caption{} \label{Fig:reg1}
    \end{subfigure}
    \vspace{0.4cm}

    \begin{subfigure}[p]{8 cm} 
        \centering
        \includegraphics[width=\textwidth]{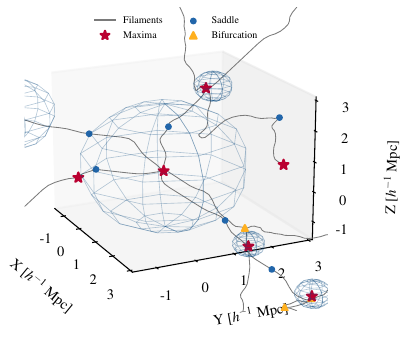} 
        \caption{} \label{Fig:reg6-zoom-center}
    \end{subfigure}
    \begin{subfigure}[p]{8 cm} 
        \centering
        \includegraphics[width=\textwidth]{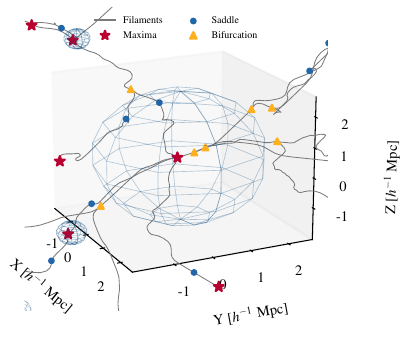} 
        \caption{} \label{Fig:reg2-zoom-bif}
    \end{subfigure}
    \caption{Examples of DisPerSE simulated networks. Panel (\subref{Fig:reg1}): Filaments in region 1 at $z=0$. The full filament network is shown in grey. The nodes are shown as red stars, the saddle points as blue dots, and the bifurcation points as yellow triangles. The light blue spheres are the $R_{200}$ of matched clusters. Panel (\subref{Fig:reg6-zoom-center}): Zoomed-in view of the central cluster of region 6. Panel (\subref{Fig:reg2-zoom-bif}): Example of two bifurcation points inside the $R_{200}$ sphere of the central cluster of region 2.}
\end{figure*}

An example of a simulated region at $z=0$ is shown in Fig. \ref{Fig:reg1}. Here, we display the whole simulated skeleton in grey, while the critical points are represented in the following way: the nodes as red stars, the saddle points as blue dots, and the bifurcation points as yellow triangles. Moreover, we plot the $R_{200}$ spheres of the matched central cluster and less massive groups in the region. A zoom-in of the central cluster and an example of bifurcation points inside the $R_{200}$ sphere are displayed in the bottom panels of the figure, respectively, in Figs. \ref{Fig:reg6-zoom-center} and \ref{Fig:reg2-zoom-bif}.

\section{Results}
\label{sec:4}
To investigate the dependence of cosmic web filaments on the properties of galaxy clusters, we estimated the connectivity of clusters, denoted with \textit{k}. The connectivity of a galaxy cluster, as it is defined by \citet{Codis:2018}, is the number of filaments globally connected with the cluster. More specifically, the connectivity is the number of filaments connected to the galaxy cluster at a specific aperture, or in other words, the number of filaments intersecting the surface of a sphere with a certain radius, $R_{\Delta}$\footnote{The subscript $\Delta$ represents the overdensity, i.e. value of the ratio between the density of the cluster inside that radius and the critical density of the Universe $\rho_c = 3H^2 / (8 \pi G)$ at the cluster's redshift.}. 

For the complete dataset, we estimated the connectivity of groups and clusters at $R_{500}$, $R_{200}$, and $R_{vir}$, denoted, respectively, as $k_{500}$, $k_{200}$, and $k_{vir}$, where the subscript $vir$ refers to the virial radius\footnote{The virial radius refers to the aperture where the halo satisfies the conditions of the virial theorem.}. With this definition, the connectivity value takes into account filaments that start from a node or a bifurcation point within the aperture considered. An example of this phenomenon is displayed in the bottom right panel of Fig \ref{Fig:reg2-zoom-bif}, where the cluster has a connectivity, $k_{200}$, equal to 7, considering the filaments originating from the bifurcation points inside $R_{200}$. We observe that, for the sake of redundancy, we omit from the connectivity symbol the subscripts denoting the cosmic web tracer and dimensionality, which in our case should be $gas$ and 3D, although these aspects should always be specified when considering the cosmic web identification. 

\subsection{Connectivity and galaxy cluster mass}
\label{sec:4.1}
The number of filaments connected to a galaxy cluster -- the connectivity -- is expected to scale with the cluster's mass. As filaments feed the cluster centre with DM, gas, and galaxies, more connected clusters are on average more massive.

\begin{figure}[h]
   \centering
   \includegraphics[width=8.8cm]{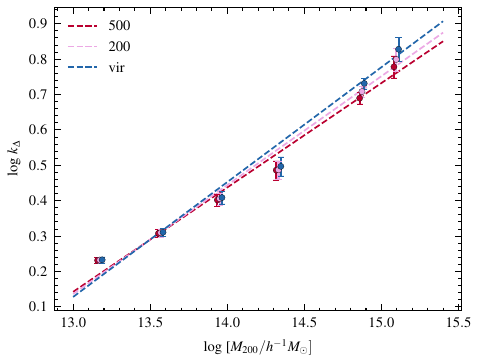}
   \caption{Mean connectivity of \textsc{The Three Hundred} groups and clusters at $z=0$ as a function of the mass, $M_{200}$. The connectivity was estimated at different over-densities: respectively, $k_{500}$ in red, $k_{200}$ in pink, and $k_{vir}$ in blue. The dashed lines represent the linear fittings. The error bars are the errors on the mean values, computed with the bootstrap method, multiplied by two for visual clarity. The mean values have been shifted in mass for visual clarity.}
              \label{Fig:k-m-delta-z0}%
\end{figure}

In \textsc{The Three Hundred} dataset, we estimated the connectivity for 3412 groups and clusters at $z=0$, with a total mass range of $10^{13} \leq M_{200} \, h^{-1} \, M_{\sun} \leq 2.62 \times 10^{15}$. Our results of the connectivity as a function of the cluster's mass, $M_{200}$, at $z=0$ are presented in Fig. \ref{Fig:k-m-delta-z0}. We find the expected correlation between mass and \textit{k} at all of the apertures explored -- $R_{500}$, $R_{200}$, and $R_{vir}$ -- whose mean values and errors on the mean, computed with the bootstrap method\footnote{For the bootstrap re-sampling, we have drawn 1000 new samples with a replacement, i.e. the same data point can be selected several times, each sample with the same length as the original sample. The error is the standard deviation of the new distribution.}, are plotted in red, pink, and blue, respectively. We performed linear fittings, plotted as dashed lines in Fig. \ref{Fig:k-m-delta-z0} and whose parameters are shown in Table \ref{Table:par-k-m}, as the following: $ \log k_{\Delta} = A \cdot \log M_{200} + B $. 

\begin{table}[h]
\caption{Fitting parameters for the $\log k_{\Delta} - \log M_{200}$ relation}             
\label{Table:par-k-m}      
\centering                          
    \begin{tabular}{c c c }        
    \hline\hline                 
    Aperture &  A & B\\    
    \hline           
    $R_{500}$ & 0.295 $\pm$ 0.016  &  $-3.69$ $\pm$ 0.24  \\ 
    $R_{200}$ & 0.308 $\pm$ 0.019 &  $-3.86$ $\pm$ 0.27  \\
    $R_{vir}$ & 0.324 $\pm$ 0.021 &  $-4.09$ $\pm$ 0.30   \\
    \hline                                   
    \end{tabular}
\end{table}

We observe a flattening in the growing trend among low-massive clusters, specifically in the range $13 \leq \log [M_{200} / h^{-1} \, M_{\sun}] \leq 14.5$, which is not present for the massive and central clusters. This phenomenon is likely attributed to the intrinsic limitation on the minimum connectivity value, introduced by our sample construction, which cannot be less than one. The clusters for which we computed the connectivity are, in fact, matched with DisPerSE critical nodes, which are, by definition, all linked to at least one filament. 

A direct computational consequence of including other critical points, besides the node matched with the cluster, inside the aperture in the estimate of the connectivity is that the connectivity depends on the aperture chosen, with $k_{vir}$ always equal to or larger than $k_{200}$ and $k_{500}$. This result is also visible in Fig. \ref{Fig:k-m-delta-z0}. We note that this effect is more evident for more massive clusters -- the clusters at the centre of the simulated boxes -- which lie in a denser environment and can have several substructures, leading to more nodes identified by DisPerSE. Further studies are needed to investigate other physical explanations of this phenomenon, such as the diffusion of filaments inside the apertures considered. This result is consistent at larger redshifts -- namely, for $z=1$ and $z=2$ -- as is shown in Appendix \ref{app:a}.  

Finally, it is worth noting that the linear fits and the findings shown in Fig. \ref{Fig:k-m-delta-z0} were computed based on a constant mass, $M_{200}$, for both the dataset at $R_{500}$ and $R_{Vir}$. This decision was made specifically to emphasise how the connectivity of a fixed cluster mass varies with the aperture. Nevertheless, it is important to mention that the positive correlation between the connectivity and mass would persist even if we were to consider the general relationship between $k_{\Delta}$ - $M_{\Delta}$.

\begin{figure}[h]
   \centering
   \includegraphics[width=8.8cm]{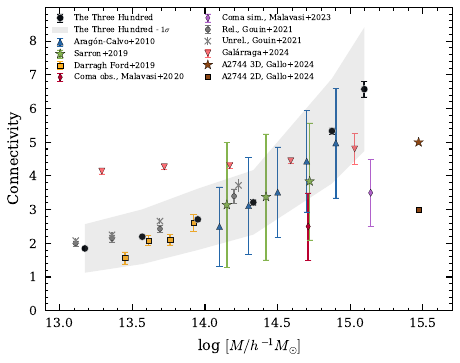}
   \caption{Connectivity of \textsc{The Three Hundred} clusters compared to literature values (see the legend for details). The black dots are the mean connectivity values, $k_{200}$, of \textsc{The Three Hundred} groups and clusters. The error bars of \textsc{The Three Hundred} points are the errors on the mean, estimated with the bootstrap method, while the shaded grey area is the standard deviation error. For the literature values, the error bars represent the error on the mean, except for the dataset of \citet{Aragon-Calvo:2010b, Sarron:2019, Malavasi:2020b} and \citet{Malavasi:2023}. For details and differences of each sample, see the main text and Table \ref{table:comparison}. }
              \label{Fig:k-m-comparison}%
\end{figure}

Our measurements of the connectivity show a strong agreement with other connectivity values from the literature, over a wider mass range. This result is shown in Fig. \ref{Fig:k-m-comparison}, where we plot \textsc{The Three Hundred} data compared to other connectivity samples from the literature, estimated from simulations and observations. 
The data of \citet{Aragon-Calvo:2010b}, plotted as blue triangles, were derived from a $\Lambda$ cold DM N-body simulation, where a multi-scale morphological filter (MMF; \citet{Aragon-Calvo:2007}) was used to extract cosmological structures such as walls, clusters, and filaments. 
The values of \citet{Gouin:2021} and \citet{Galarraga-Espinosa:2024} were estimated from hydrodynamical simulations -- the IllustrisTNG and the MilleniumTNG, respectively -- and are plotted as grey dots (relaxed groups), crosses (unrelaxed groups), and pink triangles. 
The connectivity value of the simulated Coma cluster of \citet{Malavasi:2023}, plotted as a violet diamond, was estimated from a DM-only N-body simulation. 
The observed samples are the AMASCFI cluster sample, plotted as green stars, extracted from the CFHTLS photometric redshift survey of \citet{Sarron:2019}, and the COSMOS2015 photometric redshift catalogue of \citet{DarraghFord:2019}, plotted as yellow squares.
The observed connectivity value of the Coma cluster in \citet{Malavasi:2020b} was computed from the cosmic web extracted from the Main Galaxy Sample of the Sloan Digital Sky Survey (SDSS) \citep{Malavasi:2020a} and plotted as a red diamond. 
Lastly, plotted as a brown star and square, respectively, are the 3D and 2D connectivity values of cluster Abell 2744 \citep{Gallo:2024}, computed from the spectroscopic galaxy catalogue of \citet{Owers:2011} and X-ray data observed by \textit{XMM-Newton} X-ray Observatory \citep{Eckert:2015,Jauzac:2016}.

Although the connectivity values span almost the same ranges, it is important to note the differences in how measurements were taken. This difference may include variations in how the connectivity is defined and at which aperture it is estimated, the type of tracers used to extract the cosmic web, and the filament finder employed. 

\paragraph{\textit{Aperture.}} The connectivity can be computed at a fixed aperture for all the clusters, as in \citet{Aragon-Calvo:2010b}, where they adopt a fixed aperture of $3 \, h^{-1}$ Mpc, and in \cite{Sarron:2019}, with a fixed aperture of $1.5 $ cMpc, or it can be estimated at a set overdensity, as in the case of \citet{DarraghFord:2019}, \citet{Malavasi:2020b, Malavasi:2023} and \citet{Gallo:2024}, whose connectivity is at $1.5 \, R_{vir}$, or as \citet{Gouin:2021} and \citet{Galarraga-Espinosa:2024}, with an aperture of $1.5 \, R_{200}$.


\paragraph{\textit{Web finder.}} The different web finder employed can also affect the recovery of the filamentary structures -- and therefore the estimate of the connectivity -- since they rely upon different approaches (we refer the reader to \citet{Libeskind:2018} for a review of the most common web finders). 
The most used filament finder is DisPerSE, employed in this work as well as in \citet{Sarron:2019, DarraghFord:2019, Malavasi:2020b, Malavasi:2023} and \citet{Galarraga-Espinosa:2024}, although with different parameters tuning. The cosmic web in \citet{Gouin:2021} and \citet{Gallo:2024} was extracted with the graph-based algorithm T-ReX \citep{Bonnaire:2020}.

\paragraph{\textit{Cosmic web tracer}} 
In this work, the filamentary structures are traced by the gas particle distribution, in \citet{Aragon-Calvo:2007} by DM particles, by X-ray data in the 2D analysis of \citet{Gallo:2024}, while the remaining studies make use of observed or simulated galaxy catalogues.

Lastly, we note that filaments' identification from 2D tracers can lead to an underestimation of connectivity values, as filaments along the line of sight are often only visible with 3D tracers. This effect is evident in the analysis of \citet{Gallo:2024}, where the 2D connectivity value of cluster A2744 is smaller than the 3D estimate, due to two filaments along the line of sight being identified only in the 3D analysis using a spectroscopic galaxy catalogue. It is worth noting that the connectivity values of A2744, although they are not directly comparable because of its larger mass, are still lower than those expected from the other analyses presented in Fig. \ref{Fig:k-m-comparison}. The smaller region analysed for the filament identification ($\sim4 \, \textrm{Mpc}$ radius for the 3D analysis and $\sim 3 \, \textrm{Mpc}$ radius for the 2D one) and the sparse galaxy catalogue (305 galaxies) could be the causes of, this difference.

The large difference between the trend followed by \textsc{The Three Hundred} points and the measurements by \citet{Galarraga-Espinosa:2024} at masses lower than $10^{14} \, h^{-1} \, M_{\odot}$ may be related to different choices of the persistence threshold. In general, lower persistence values are needed to connect the extensive sample of clusters of \citet{Galarraga-Espinosa:2024}, which spans a mass range from $10^{11}$ to $10^{15} \, h^{-1} \, M_{\odot}$, not entirely shown in Fig. \ref{Fig:k-m-comparison}. Lower persistence values yield higher numbers of filaments, including weaker and smaller structures, increasing the connectivity measure. A second possible explanation for this difference is suggested in \citet{Boldrini:2024}. They show that a similar trend to the one of \citet{Galarraga-Espinosa:2024} emerges when the analysis solely includes groups and clusters with a minimum connectivity of 2. When imposing the same threshold in connectivity -- excluding from the analysis the groups with k = 1 -- we do observe a small increase in the average connectivity measurements of groups less massive than $10^{14} \, h^{-1} \, M_{\odot}$, although this is still insufficient to match the values of \citet{Galarraga-Espinosa:2024}. 


We did not include in our study, or in Fig. \ref{Fig:k-m-comparison}, the data published in \citet{Einasto:2021} and \citet{Aghanim:2024} of clusters in the Corona Borealis supercluster and the Shapley supercluster, respectively. These clusters do not occupy the same parameter space as \textsc{The Three Hundred} clusters and are not directly comparable, as their connectivity measurement may be biased by the dense environment in which they reside. A future investigation into the connectivity and mass relation for clusters within superclusters could be crucial for understanding the impact of a cluster's environment on its connectivity measurement. 

All the studies presented in Fig. \ref{Fig:k-m-comparison} confirm a positive correlation between the number of filaments connected to a cluster and its mass, affirming the significant role filaments play in the accretion of matter towards the clusters. However, it is important to take the possible biases into account, summarised in Table \ref{table:comparison}, when referring to the connectivity.

\begin{table*}[t]
\caption{Parameters of previous literature studies compared to this work.}             
\label{table:comparison}      
\centering                          
    \begin{tabular}{c c c c c c c}        
    \hline\hline                 
    Sample & Data & $k_R$ & Mass & CW extraction & Web finder & References \\    
    \hline           
    The300 & Hydro-sim & $R_{200}$ & $M_{200}$ & 3D gas particles & DisPerSE & this work \\ 
    N-body sim & DM-sim & $3 \, h^{-1}$ Mpc & $~M_{vir}$ & 3D DM particles & MMF & 1 \\ 
    AMASCFI & Observations & $1.5 \, h^{-1}$ Mpc & $M_{200}$ & 2D galaxies & DisPerSE & 2 \\ 
    COSMOS2015 & Observations & $1.5 \, R_{vir}$ & $M_{200}$ & 2D galaxies & DisPerSE &  3 \\ 
    Coma obs. & Observations & $1.5 \, R_{vir}$ & $M_{200}$ & 3D galaxies & DisPerSE & 4,5 \\ 
    Coma sim. & DM-sim & $1.5 \, R_{vir}$ & $M_{200}$ & 2D galaxies & DisPerSE & 6 \\ 
    IllustrisTNG & Hydro-sim & $1.5 \, R_{200}$ & $M_{200}$& 3D galaxies & T-ReX & 7 \\ 
    MilleniumTNG & Hydro-sim & $1.5 \, R_{200}$ & $M_{200}$ & 3D galaxies & DisPerSE & 8 \\ 
    Abell 2744 (3D) & Observations & $1.5 \, R_{vir}$ & $M_{200}$ & 3D galaxies & T-Rex & 9 \\
    Abell 2744 (2D) & Observations & $1.5 \, R_{vir}$ & $M_{200}$ & X-ray map & T-Rex & 9 \\
    
    \hline                                   
    \end{tabular}

    \tablebib{
(1) \citet{Aragon-Calvo:2010b}; (2) \citet{Sarron:2019}; (3) \citet{DarraghFord:2019}; (4) \citet{Malavasi:2020b} (5) \citet{Malavasi:2020a};
(6) \citet{Malavasi:2023}; (7) \citet{Gouin:2021}; (8) \citet{Galarraga-Espinosa:2024}; (9) \citet{Gallo:2024}.
}
\end{table*}


\subsection{Connectivity evolution with redshift}

According to the standard theory of anisotropic collapse \citep{Zel'dovich:1970}, matter in the Universe is predicted to flow from less dense to denser regions over cosmic time, progressing from walls to filaments and eventually to clusters.
As a consequence, in recent times the Universe has predominantly been composed of large under-dense regions -- voids -- in terms of volume, while clusters and filaments dominate in terms of mass. 
Moreover, it has been shown (see e.g. \citet{Cautun:2014}) that, at high redshift, the filaments of the cosmic web are dominated by thin structures, which are then overtaken by more significant ones at later times.
In this framework, we investigated the evolution of cluster connectivity with redshift, which is expected to decrease with cosmic time, as is shown in \citet{Codis:2018} and \citet{Galarraga-Espinosa:2024}. 

To detect the cosmic web skeletons at higher redshifts, we created the density maps from the gas particle distributions at redshifts $z = 1.03$ and $z= 2.02$, with the same parameters used for the $z=0$ sample, with a pixel resolution of $150 \, h^{-1}$ kpc and a Gaussian smoothing with an FWHM of 4 pixels. 
We used the same DisPerSE functions and persistence cut of $0.5$, in order to extract filamentary structures with the same robustness and significance. 
The two final samples have a completeness value of $82\% $ and $83\%$, respectively, at redshifts $z=1$ and $z=2$, indicating that almost every cluster is matched by a node, similar to the sample at $z=0$. 

\begin{figure}[h]   
    \centering
   \includegraphics[width=8.8cm]{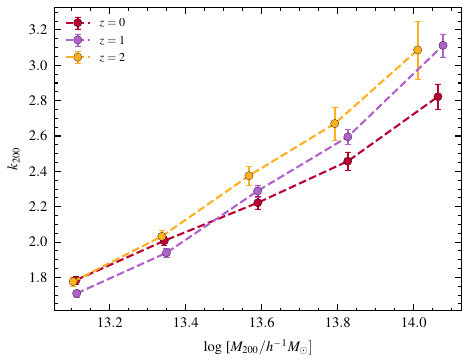}
   \caption{Evolution with redshift of the connectivity-mass relation. In yellow are shown the mean values of connectivity. $k_{200}$, extracted from the gas skeleton at $z=2$, in purple the values at $z=1$, and in red the values at $z=0$. The error bars represent the errors on the mean, computed with the bootstrap method. }
              \label{Fig:k-m-redshift}%
\end{figure}

We compared the connectivity mean values estimated at $R_{200}$ of clusters in the mass range of $10^{13} \leq M_{200} \, h^{-1} \, M_{\odot} \leq 10^{14}$, populated at all the redshifts explored. 
In Fig. \ref{Fig:k-m-redshift}, we plot the connectivity values at $z=0$ in red, $z=1$ in purple, and $z=2$ in yellow. 
We recovered the expected trend of the connectivity decrease from high redshift to present times, in agreement with what has been found both in theoretical predictions from 2D and 3D estimates \citep{Codis:2018} and, more recently, in hydrodynamical simulations \citep{Galarraga-Espinosa:2024}, although with different connectivity values, probably for a similar reason to the one discussed in Subsect. \ref{sec:4.1}. 
We observe that for masses below $2.5 \times 10^{13} \, h^{-1} \, M_{\odot}$, the connectivity at $z=0$ shows a flattening trend and remains comparable to the values of higher redshifts. 
A possible reason for this behaviour can be found in the intrinsic minimum connectivity value introduced by the sample construction, as is explained in detail in Subsect. \ref{sec:4.1}. 

The decrease of connectivity with cosmic time is both a consequence of the anisotropic collapse of the cosmic web structures, with two filaments that merge into one, or a result of their stretch and disconnection caused by the cosmological expansion of void regions in later times due to dark energy. However, further research is required to explore the influence of matter inflow from walls to filaments on filamentary evolution with cosmic time. 
We note that future studies of the evolution of connectivity with redshift with different dark energy models \citep[see e.g.][]{Codis:2018}, and critical events, such as filaments mergers \citep[see e.g.][]{Cadiou:2020}, could offer an interesting probe for dark energy models.

\subsection{Connectivity and galaxy cluster dynamical state}
\label{sec:dyn}
To investigate if the filaments are related to the dynamical state of galaxy clusters or if the connectivity can be used as a tracer of the dynamical state, we studied the correlation between the connectivity and the dynamical state of each cluster. 
In this work, we assess the dynamical state of a cluster following the definition of \citet{DeLuca:2021}, based on 3D dynamical parameters estimated from the physical properties of each particle in the simulation.
To evaluate the dynamical state in regions different from the virial radius volumes, we considered two indicators of the cluster's dynamical state, the total sub-halo mass fraction, $f_s$, and the centre-of-mass offset, $\Delta_r$. 
The total sub-halo mass fraction, $f_s$, was defined as the sum of all the sub-halo masses, $M_i$, identified by \textsc{AHF} within a specific aperture, $R_\Delta$, normalised by the total cluster mass, $M_\Delta$: 
\begin{equation}
     f_s = \frac{\sum_i M_i}{M_{\Delta}}
    \label{eq: fs deluca}
.\end{equation}
The centre-of-mass offset, $\Delta_r$, was defined as
\begin{equation}
        \Delta_r = \frac{|\textbf{r}_{cm} - \textbf{r}_c|}{R_{\Delta}}
    \label{eq: deltar}
,\end{equation}
where $\textbf{r}_{cm}$ is the position of the centre-of-mass and $\textbf{r}_{c}$ is the theoretical centre of the cluster, defined as the highest density peak, which typically coincides with the brightest cluster galaxy. The $\Delta_r$ parameter is used to describe the cluster's deviation from smoothness and spherical symmetry, as objects with high values of $\Delta_r$ are considered morphologically disturbed. 

Specifically, we have classified the clusters as ‘relaxed’ if both the conditions $f_s < 0.1$ and $\Delta_r < 0.1$ are simultaneously satisfied. On the other hand, a cluster is classified as dynamically ‘disturbed’ if both $f_s > 0.1$ and $\Delta_r > 0.1$ are true. The clusters whose indicators are not simultaneously verified -- therefore, if $f_s > 0.1$ and $\Delta_r < 0.1$ or if $f_s < 0.1$ and $\Delta_r > 0.1$ -- are classified as ‘hybrid’. In our sample at redshift $z=0$, $66\%$ of the clusters are classified as relaxed, $20\%$ as hybrid, and $14\%$ as disturbed.

\begin{figure}[h!]
    \centering
    \includegraphics[width=8.8cm]{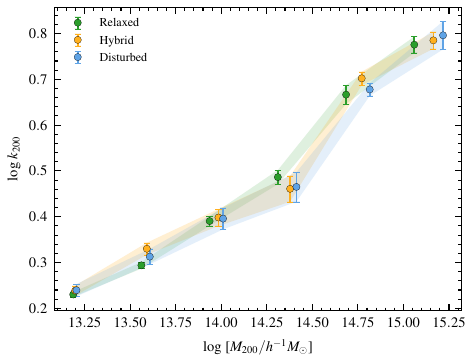}
    \caption{Connectivity and mass relation for three dynamical state sub-samples at redshift $z=0$. In green are shown the connectivity mean values, $k_{200}$, for the dynamically relaxed clusters, in yellow the values for hybrid clusters, and in light blue the ones for dynamically disturbed clusters. The error bars represent the errors of the mean, computed with the bootstrap method. }
    \label{Fig:k-m-chi-z0}%
\end{figure}

In Fig. \ref{Fig:k-m-chi-z0}, we plot the connectivity-mass relation, computed at $R_{200}$, for the three sub-samples of dynamical state; respectively, relaxed clusters in green, hybrid ones in yellow, and unrelaxed ones in light blue. The three sub-samples overlap and are comparable to each other, within the errors estimated with the bootstrap method, indicating that the connectivity-mass relation is not affected by the dynamical state -- assessed in terms of the two indicators, $f_s$ and $\Delta_r$ -- of the cluster. This result is not dependent on the redshift, as is shown in Appendix \ref{app:b}, where we present this analysis at redshifts $z=1$ and $2$, or on the overdensity at which the connectivity and the dynamical state are estimated, as is shown in Appendix \ref{app:c}, where we present the same analysis for $R_{500}$ and $R_{vir}$.

To further investigate the relation between the dynamical state of the cluster and the filaments connected to it, we now focus on the continuous dynamical state indicator, $\chi$. The degree of relaxation combines the dynamical parameters in a single measure of the cluster dynamical state and is defined by \citet{Haggar:2020} as 
\begin{equation}
    \chi = \left[\frac{\sum_i (\frac{x_i}{x_{0,i}})^2 }{N} \right]^{-1/2}
    \label{eq:degreerel}
,\end{equation}
where $x_i$ is the general indicator of the dynamical state, $x_{0,i}$ is the threshold between the relaxed and the disturbed regimes, and $N$ is the number of indicators used. As a result, larger values of $\chi$ (with $\log \chi > 0$) indicate a more relaxed dynamical state, while negative values of $\log \chi$ represent a disturbed dynamical state.
In the literature, the definition of the degree of relaxation can vary both in the choice of the parameters used and in the selection of the threshold values \citep[see e.g.][]{Cui:2017}. In this work, we estimated $\chi$ with $f_s$ and $\Delta_r$, both considered with a threshold of $0.1$. We estimated the dynamical state of the clusters at redshift $z=0$ and computed the degree of relaxation as in Eq. \ref{eq:degreerel}, considering the parameters at $R_{200}$. 

To possibly disentangle the impact of connectivity and mass on the dynamical state, given their correlation, we divided the total dataset into three bins of connectivity and three bins of mass. This approach is necessary to study the relationship between the degree of relaxation and the connectivity independently of the mass, and vice versa. 

The three connectivity bins divide the dataset into $\sim$700 weakly connected groups with $k_{200} < 2$, $\sim$2000 average-connected clusters with $k_{200} \in \{ 2,3 \}$, and $\sim$500 highly connected clusters with $k_{200} > 3$. These thresholds were chosen taking the mean connectivity value of $k_{200} = 2.42 $ of the whole cluster population at redshift $z=0$ into account.

\begin{figure*}[h]
   \centering
   \begin{subfigure}[h]{8.8cm}
       \centering
       \includegraphics[width=\textwidth]{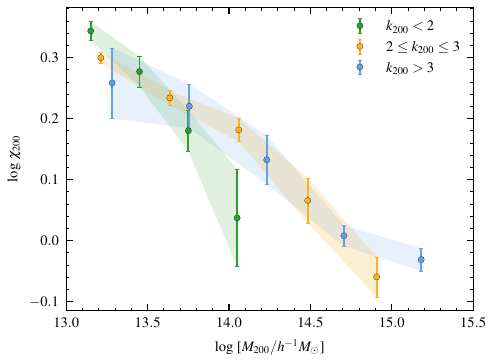}
        \caption{}
        \label{Fig:chi-m-ksep-z0}%
   \end{subfigure}
   \begin{subfigure}[h]{8.8cm}
       \centering
       \includegraphics[width=\textwidth]{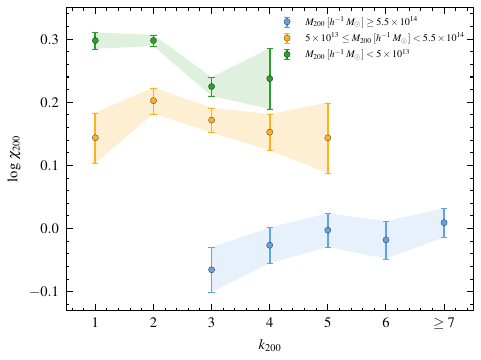}
        \caption{}
        \label{Fig:chi-k-masssep-z0}%
   \end{subfigure}

   \caption{Dynamical state dependence on connectivity and cluster total mass at redshift $z=0$. Panel (a): Degree of relaxation, $\chi_{200}$, as a function of the total mass, $M_{200}$. Green, yellow, and light blue dots indicate, respectively, weakly connected ($k_{200} < 2$), average-connected ($k_{200} \in \{ 2,3 \}$), and highly connected ($k_{200} > 3$) clusters. Panel (b): Degree of relaxation, $\chi_{200}$, as a function of the connectivity, $k_{200}$. Green, yellow, and light blue dots indicate, respectively, low-mass ($M_{200} < 5 \times 10^{13} \, h^{-1} \, M_{\odot}$), medium ($5 \times 10^{13} \leq M_{200} \, h^{-1} \, M_{\odot} < 5.5 \times 10^{14} $), and massive ($M_{200} \geq 5.5 \times 10^{14} \,  h^{-1} \, M_{\odot}$) clusters. The shaded areas and the error bars indicate the error of the mean, computed with the bootstrap method.}
   \label{Fig:dynstate-general}
   
\end{figure*}

In the left panel (a) of Fig. \ref{Fig:dynstate-general}, we plot the mean degree of relaxation as a function of mass for the three connectivity sub-samples; respectively, in green the weakly connected clusters, in yellow the average-connected ones, and in light blue the highly connected ones. At fixed mass, the three connectivity bins overlap and are comparable to each other within the errors, confirming that the connectivity does not influence the dynamical state of the cluster, computed with the $f_s$ and $\Delta_r$, corroborating the result shown in Fig. \ref{Fig:k-m-chi-z0}.

This outcome is not consistent with that of \citet{Gouin:2021} who found a correlation between the degree of relaxation and the connectivity values, with more connected clusters being on average more dynamically unrelaxed. We note that the disagreement could be due to the different definition of the degree of relaxation used in \citet{Gouin:2021}, which differs from ours in the dynamical indicators and thresholds used, as well as the differences in the dataset and the cosmic web extraction, as is illustrated in Table \ref{table:comparison}. 

Focusing now on the mass bins, we divided the dataset into low-mass groups with $M_{200} < 5 \times 10^{13} \, h^{-1} \, M_{\odot}$, medium groups and clusters with $ 5 \times 10^{13} \leq M_{200} \, h^{-1} \, M_{\odot} < 5.5 \times 10^{14} $, and massive clusters with $M_{200} \geq 5.5 \times 10^{14} \,  h^{-1} \, M_{\odot}$, which correspond to the large galaxy clusters at the centre of each simulated region.
The three sub-samples are plotted in the right panel (b) of Fig. \ref{Fig:dynstate-general} in green, yellow, and light blue, respectively. Contrary to the connectivity sub-samples shown in Fig. \ref{Fig:chi-m-ksep-z0}, at fixed connectivity the three mass bins show a distinct separation from one another. This behaviour suggests a correlation between the degree of relaxation, $\chi_{200}$, and the mass of the cluster, independently of their connectivity value, with less massive groups appearing to be more dynamically relaxed, and vice versa. 

This result was also reported by \citet{Kuchner:2020} -- although with a different definition of the degree of relaxation, both in the parameters and the thresholds chosen -- who suggest that the most massive clusters tend to be in a more dynamically disturbed state. This phenomenon is likely attributed to the fact that the most massive clusters are currently still growing, resulting in larger centre-of-mass offsets and in the presence of internal substructures \citep{Power:2012}. Moreover, the high-density environment in which they are situated can make them more likely to accrete matter, resulting in less relaxed states \citep{Kuchner:2020}.

\noindent

\subsection{Connectivity and mass bias}

Classical approaches to inferring the mass of galaxy clusters from SZ and X-rays observations usually rely on the hydrostatic equilibrium (HE) assumption \citep[for a review, see][]{Kravtsov:2012}. This hypothesis assumes that the gas thermal pressure is balanced by the gravitational forces, leading the cluster to a state of equilibrium. Deviations from the HE can introduce errors in the estimation of observable properties of clusters and their mass. 

We investigated if the presence of filaments, quantified with the connectivity, has an impact on the HE of the galaxy clusters of \textsc{The Three Hundred} simulation. For this purpose, we refer to the analysis of \citet{Gianfagna:2023}, who explored the deviations of the HE mass, $M_{HE}$, from the total mass $M_{true}$ of the 324 simulated central clusters, quantified with the hydrostatic mass bias, $b$. 
$M_{true}$ refers to the real mass, computed by summing all the DM, gas, and star particle masses inside a radius. On the other hand, the HE mass, $M_{HE}$, is the mass inferred from the gas thermal pressure profile estimated from SZ observations or from the temperature and electron density profiles from X-ray observations, leading, respectively, to the mass biases $b_{SZ}$ and $b_{X}$, defined as: 

\begin{equation}
    b = \frac{M_{true} - M_{HE}}{M_{true}}
.\end{equation}
We refer to \citet{Gianfagna:2023} for the complete derivation of the parameters. We note that a positive bias, if defined in this way, denotes an underestimation of the true mass. 

\begin{figure}[h!]
    \centering
    \includegraphics[width=8.8cm]{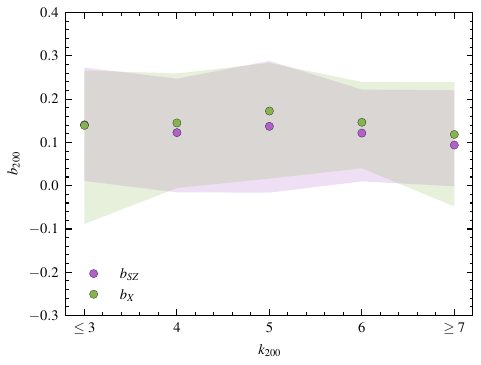}
    \caption{Connectivity and hydrostatic mass bias relation at redshift $z=0$. In green are shown the mass bias median values, $b_{X}$, at $R_{200}$  and in violet the mass bias median values, $b_{SZ}$, at $R_{200}$. The shaded areas represent the 16th and 84th percentiles. }
    \label{Fig:k-bias-200}%
\end{figure}

We plot in Fig. \ref{Fig:k-bias-200} the median values of the hydrostatic mass biases, $b_{SZ}$ and $b_{X}$, estimated at $R_{200}$ as a function of the connectivity, $k_{200}$, of the cluster. No correlation is found between the bias and the connectivity, independently of the observable property used to estimate it. 
Moreover, we note that we do not observe any variation in the scatter with connectivity, meaning that the dispersion of the bias is independent of the number of filaments connected. 
On the other hand, the scatter of the hydrostatic mass bias has been found to vary as a function of the dynamical state, quantified with different indicators, with disturbed clusters showing a wider dispersion \citep{Piffaretti:2008,Rasia:2012,Ansarifard:2020,Gianfagna:2023}. This result suggests that the connectivity may not be used as a reliable indicator of the dynamical state of a galaxy cluster, as was already shown in Sec. \ref{sec:dyn}. 
The same result is observed for the connectivity and the bias estimated at $R_{500}$, as is shown in Appendix \ref{app:d}.

\section{Mock-galaxy filaments}
\label{sec:5}
In this section, we aim to investigate the effect of the chosen tracer on the extraction of the cosmic web. In this context, we compare the connectivity estimates from filaments extracted from the gas particle distribution to those identified from the mock-galaxy distribution in \textsc{The Three Hundred} regions. 

The simulated cosmic web is traced by the mock galaxy catalogue of the full $1 \, h^{-1}$ Gpc MDPL2 box at redshift $z=0$. Haloes are defined as a mock galaxy if they have a minimum stellar mass of $M_* \geq 3 \times 10^9 \, h^{-1} \, M_{\odot}$, which corresponds to haloes with $M_{halo} > 3 \times 10^{10} \, h^{-1} \, M_{\odot}$. The density is computed from the Delaunay tessellation of the points with the DTFE.
The tessellation is mass-weighted, with a weight associated with each tessellation vertex corresponding to the mass of the halo at this vertex, to improve the filament identification, as is explained in detail by \citet{Kuchner:2020}. The cosmic web skeleton is then identified with DisPerSE with a persistence ratio threshold of $6.5 \sigma$ 
, which provides the best agreement with the cosmic web extracted from gas particles, as is tested in \citet{Kuchner:2020}. A more detailed description of the extraction procedure and in-depth analysis of filament properties from the full MDPL2 box will be provided in a dedicated paper (Bianconi et al., in prep). 

We associated the DisPerSE nodes with simulated groups, considering as matched those with a distance of less than 4 pixels, as is explained in detail in Sect. \ref{sec:3.2}. The final sample, for all the 324 regions, includes 836 galaxy groups and clusters with a mass range of $10^{13} \leq M_{200} \, h^{-1} \, M_{\sun} \leq 2.69 \times 10^{15}$. Lastly, we estimated the connectivity, $k_{200}$, of the galaxy clusters at $R_{200}$, as is defined in Sect. \ref{sec:4}. 

\begin{figure}[h!]
    \centering
    \includegraphics[width=8.8cm]{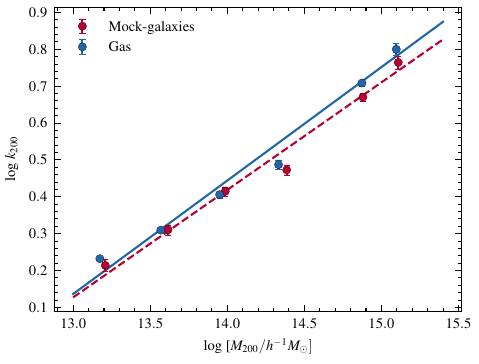}
    \caption{Mean connectivity of \textsc{The Three Hundred} groups and clusters at $z=0$, estimated from gas and mock-galaxy filaments. The mean values of connectivity, $k_{200}$, from gas filaments are plotted in blue, and the values of $k_{200}$ from mock-galaxy filaments are plotted in red. The solid and dashed lines represent the linear fitting for the gas and the mock galaxies, respectively. The error bars are the errors on the mean values, computed with the bootstrap method.}
    \label{Fig:k-m-gal}%
\end{figure}

The results of the connectivity, $k_{200}$, estimated from mock-galaxy filaments are plotted in Fig. \ref{Fig:k-m-gal} in red, as a function of the cluster's mass, $M_{200}$. We plot as blue dots the connectivity values computed from gas filaments, presented in Sect. \ref{sec:4.1}. We performed the same linear fitting as for the gas connectivity, whose parameters are shown in Table \ref{Table:par-k-m-galaxy}.

\begin{table}[h]
\caption{Fitting parameters for the $\log k_{200} - \log M_{200}$ relation for gas-traced filaments and mock-galaxy-traced filaments. }             
\label{Table:par-k-m-galaxy}      
\centering                          
    \begin{tabular}{c c c }        
    \hline\hline                 
    Tracer &  A & B\\    
    \hline           
    Mock-Galaxies & 0.291 $\pm$ 0.018  & $-3.66
    $ $\pm$ 0.26 \\ 
    Gas & 0.308 $\pm$ 0.019 &  $-3.86$ $\pm$ 0.27  \\
    \hline                                   
    \end{tabular}
\end{table}

The strong agreement between the two connectivities is an indication that the number of filaments traced by gas or mock galaxies is similar, and therefore that these are representative of the same structures. A similar result in \textsc{The Three Hundred} simulation was shown in \citet{Kuchner:2020}, where they measure the distance between gas-traced and galaxy-traced filaments. They find a median value of $0.67 \, h^{-1} \, Mpc$ for the distribution of distance between the two skeletons. This value lowers to $0.47 \, h^{-1} \, Mpc$ when taking only mass-weighted galaxies into account, as in the case of our work. Both values are lower than the typical thickness of a filament. However, potential discrepancies between the skeletons extracted from gas or DM particles and haloes catalogues are expected, as has been noted in other studies \citep[e.g.][]{Bond:2010a,Bond:2010b,Laigle:2018,Kuchner:2020, Zakharova:2023}. These discrepancies are likely to be attributed to the finer sampling of the density field by gas particles, which outnumber the mock galaxies, and therefore could result in reduced noise during the cosmic web extraction process. Moreover, employing different density field estimation methods -- a regular grid for the gas particles and the Delaunay tessellation for the mock galaxies -- may introduce some variations in the extracted skeletons, such as a preferred direction for the filamentary structures. The discrepancies between the two tracers and the effects on the recovered cosmic web are investigated more in detail by Bianconi et al. (in prep).

\section{Conclusions}
\label{sec:6}
The outskirts of galaxy clusters can provide insightful information for the study of large-scale structures. In particular, the presence of filamentary structures, acting as pathways for matter transportation towards the centre of clusters, and their impact on clusters' properties can be investigated. The aim of this study is to examine the correlation between the connectivity and the main galaxy clusters properties.  

In this work, we have used the 324 regions with a $30 \, h^{-1}$ Mpc side of \textsc{The Three Hundred} hydrodynamical simulation to extract the filamentary structures around galaxy clusters and groups. We ran the DisPerSE algorithm on the gas particle density maps at redshifts $z=0,1$, and $2$. The full procedure of generating the density maps has been described, together with the setting of relevant parameters of DisPerSE. We have estimated the connectivity at several apertures, in particular at $R_{500}$, $R_{200}$, and $R_{vir}$ for $\sim 10^3$ groups and clusters with a total mass range of $10^{13} \leq M_{200} \, h^{-1} \, M_{\sun} \leq 2.62 \times 10^{15}$ at redshift $z=0$. \\
The main results of this work can be summarised as follows: 
\begin{enumerate}

    \item[(i)] The connectivity, computed as the number of gas filaments crossing the sphere at a specific aperture, quantified in terms of overdensity, depends on the chosen radius, with larger values of connectivity when estimated at larger radii. This is a consequence of including filaments that start from substructures -- maxima and bifurcation points -- that lie within the radius considered. Therefore, it is important to specify the aperture chosen when analysing the connectivity of galaxy clusters. 
    \item[(ii)] Filaments have an impact on galaxy cluster mass accretion, transporting matter from the outskirts to the centre. On average, more massive galaxy clusters are also more connected. This result is in agreement with several literature studies, both from observed and simulated datasets. 
    \item[(iii)] We observe a decrease in the connectivity mean values over cosmic time, with clusters being more connected at higher redshifts. This result is likely to be a consequence of the anisotropic collapse of the cosmic web structures.
    \item[(iv)] No dependence of connectivity with the cluster's dynamical state is found. The connectivity-mass correlation does not have any significant difference when computed for three sub-samples of clusters: dynamically relaxed, hybrid, and disturbed. Moreover, we do not observe any correlation between the connectivity and the degree of relaxation parameter, $\chi$. On the other hand, we find that, independently of the connectivity, more massive clusters are on average more dynamically disturbed.
    \item[(v)] No correlation of the hydrostatic mass bias with redshift is found for the 324 central clusters. This result is independent of the aperture at which the quantities are computed and the observable properties used to infer the hydrostatic mass bias. Moreover, we do not observe a variation in the scatter of the bias with the connectivity, contrary to what is found with the dynamical state of clusters. This suggests that the connectivity may not be an indicator of the dynamical state.
    \item[(vi)] We observe a strong agreement between the connectivity evaluated from gas-traced filaments and mock-galaxy-traced filaments, extracted from the mock haloes catalogue of the full MDPL2 box. This result suggests that gas and mock galaxies trace a comparable number of structures, yielding a similar connectivity value, and offer similar conclusions on the overall connections of the cosmic web.
\end{enumerate} 

This work underlines the dependence of filaments, traced by the gas particle distribution, on the properties of the galaxy clusters to which they are connected and shows a correlation between the connectivity and the total mass of the clusters. This result affirms the role that filaments play in the accretion of matter towards the centre of the cluster. 

The results presented in this work and the simulated cosmic web skeletons will be used as a reference for cosmic web extraction from Compton-y simulated maps, to provide insightful forecasts for observations. 
With further investigations on filaments' evolution with cosmic time, useful information on the study of cosmology and structure evolution can be provided. 
Further work is needed to assess the differences in the cosmic web extraction process from distinct tracers and from different density field estimation methods. Future analysis with different hydrodynamical codes, such as the \textsc{Gizmo-Simba} run of \textsc{The Three Hundred} \citep{Cui:2022}, will be necessary to investigate the dependence of the results on different baryonic models.


\begin{acknowledgements}
The authors would like to thank the referee for the helpful comments and suggestions. This work has been made possible by \textsc{The Three Hundred} collaboration. The simulations used in this paper have been performed in the MareNostrum Supercomputer at the Barcelona Supercomputing Center, thanks to CPU time granted by the Red Española de Supercomputación. As part of \textsc{The Three Hundred} project, this work has received financial support from the European Union’s Horizon 2020 Research and Innovation programme under the Marie Skłodowska-Curie grant agreement number 734374, the LACEGAL project. S.S. acknowledges support from Sapienza Università di Roma thanks to "Progetti per Avvio alla Ricerca", n. AR123188AF0D68BF, and "Bando Mobilità Internazionale PhD", D.R. n. 1505/2023. M.D.P., S.S. and A.F. acknowledge financial support from PRIN 2022 (Mass and selection biases of galaxy clusters: a multi-probe approach - n. 20228B938N) and from Sapienza Università di Roma, thanks to Progetti di Ricerca Medi 2022, RM1221816758ED4E. G.Y., W.C. and S.S. would like to thank also  Ministerio de Ciencia e Innovación (Spain) for financial support under project grant PID2021-122603NB-C21. A.F. acknowledges the project "Strengthening the Italian Leadership in ELT and SKA (STILES)", proposal nr. IR0000034 funded under the program "Next Generation EU" of the European Union, “Piano Nazionale di Ripresa e Resilienza” (PNRR) of the Italian Ministry of University and Research (MUR), “Fund for the creation of an integrated system of research and innovation infrastructures”, Action 3.1.1 "Creation of new IR or strengthening of existing IR involved in the Horizon Europe Scientific Excellence objectives and the establishment of networks”.

W.C. is also supported by the Atracci\'{o}n de Talento Contract no. 2020-T1/TIC-19882 granted by the Comunidad de Madrid in Spain, and the science research grants were from the China Manned Space Project.
\end{acknowledgements}

%
%
\bibliographystyle{aa}
\bibliography{export-bibtex}

\begin{appendix}
\section{Connectivity and cluster overdensity at higher redshifts}
\label{app:a}

\begin{figure}[h]
   \centering
   \includegraphics[width=8.8cm]{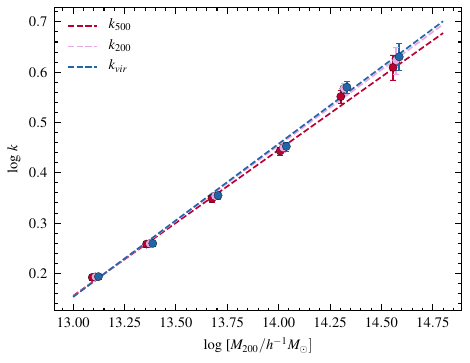}
   \caption{Mean connectivity of \textsc{The Three Hundred} groups and clusters at $z=1$ as a function of the mass $M_{200}$. The connectivity was estimated at different over-densities: respectively, $k_{500}$ in red, $k_{200}$ in pink, and $k_{vir}$ in blue. The dashed lines represent the linear fitting. The error bars are the errors on the mean values, computed with the bootstrap method. The mean values have been shifted in mass for visual clarity.}
              \label{Fig:k-m-delta-z1}%
\end{figure}

\begin{figure}[h]
   \centering
   \includegraphics[width=8.8cm]{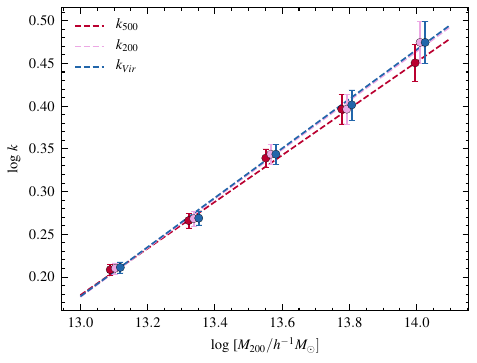}
   \caption{Mean connectivity of \textsc{The Three Hundred} groups and clusters at $z=2$ as a function of the mass $M_{200}$. The connectivity was estimated at different over-densities: respectively,$k_{500}$ in red, $k_{200}$ in pink, and $k_{vir}$ in blue. The dashed lines represent the linear fitting. The error bars are the errors on the mean values, computed with the bootstrap method. The mean values have been shifted in mass for visual clarity.}
              \label{Fig:k-m-delta-z2}%
\end{figure}

The connectivity depends on the aperture at which it is estimated, independently of the redshift. We plot the mean values of $k_{500}$, $k_{200}$, and $k_{vir}$ as a function of the cluster mass at redshift $z=1$ and $z=2$, respectively in Figs. \ref{Fig:k-m-delta-z1} and \ref{Fig:k-m-delta-z2}.

\section{Connectivity and dynamical state at higher redshifts}
\label{app:b}
\begin{figure}[h!]
    \centering
    \includegraphics[width=8.8cm]{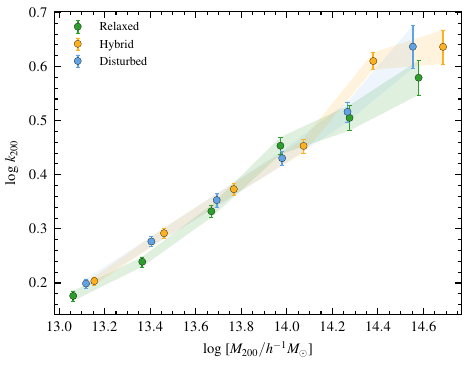}
    \caption{Connectivity and mass relation for three dynamical state sub-samples at redshift $z=1$. In green are shown the connectivity mean values, $k_{200}$, for the dynamically relaxed clusters, in yellow the values for hybrid clusters, and in light blue the ones for dynamically disturbed clusters. The error bars represent the errors of the mean, computed with the bootstrap method. }
    \label{Fig:k-m-chi-z1}%
\end{figure}

\begin{figure}[h!]
    \centering
    \includegraphics[width=8.8cm]{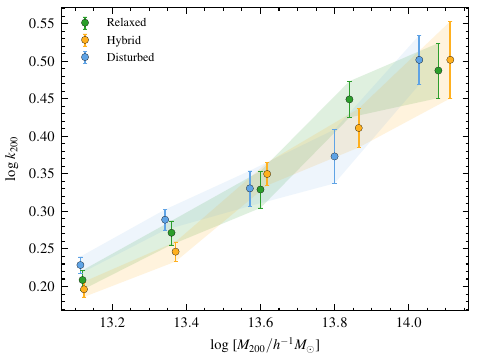}
    \caption{Connectivity and mass relation for three dynamical state sub-samples at redshift $z=2$. In green are shown the connectivity mean values, $k_{200}$, for the dynamically relaxed clusters, in yellow the values for hybrid clusters, and in light blue the ones for dynamically disturbed clusters. The error bars represent the errors of the mean, computed with the bootstrap method. }
    \label{Fig:k-m-chi-z2}%
\end{figure}

The connectivity and mass correlation is not dependent on the dynamical state of the cluster, regardless of the redshift. We plot in Figs. \ref{Fig:k-m-chi-z1} and \ref{Fig:k-m-chi-z2} the connectivity mean values as a function of the mass for dynamically relaxed, hybrid, and disturbed clusters, respectively at redshift $z=1$ and $z=2$.

\section{Connectivity and dynamical state at different overdensities}
\label{app:c}
\begin{figure}[h!]
    \centering
    \includegraphics[width=8.8cm]{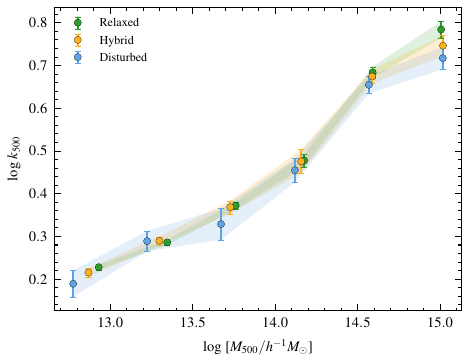}
    \caption{Connectivity and mass relation for three dynamical state sub-samples at redshift $z=0$, estimated at $R_{500}$. In green are shown the connectivity mean values, $k_{500}$, for the dynamically relaxed clusters, in yellow the values for hybrid clusters, and in light blue the ones for dynamically disturbed clusters. The error bars represent the errors of the mean, computed with the bootstrap method.}
    \label{Fig:k-m-chi-z0-r500}%
\end{figure}

\begin{figure}[h!]
    \centering
    \includegraphics[width=8.8cm]{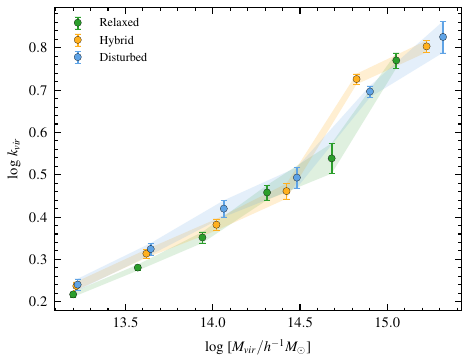}
    \caption{Connectivity and mass relation for three dynamical state sub-samples at redshift $z=0$, estimated at $R_{vir}$. In green are shown the connectivity mean values, $k_{vir}$, for the dynamically relaxed clusters, in yellow the values for hybrid clusters, and in light blue the ones for dynamically disturbed clusters. The error bars represent the errors of the mean, computed with the bootstrap method. }
    \label{Fig:k-m-chi-z0-rvir}%
\end{figure}

The connectivity and mass correlation is not dependent on the dynamical state of the cluster, independently of the overdensity at which the connectivity and the dynamical state are estimated. We plot in Figs. \ref{Fig:k-m-chi-z0-r500} and \ref{Fig:k-m-chi-z0-rvir} the connectivity mean values as a function of the mass for dynamically relaxed, hybrid and disturbed clusters, estimated respectively at $R_{500}$ and $R_{vir}$.

\section{Connectivity and HE bias at $R_{500}$}
\label{app:d}

\begin{figure}[h!]
    \centering
    \includegraphics[width=8.8cm]{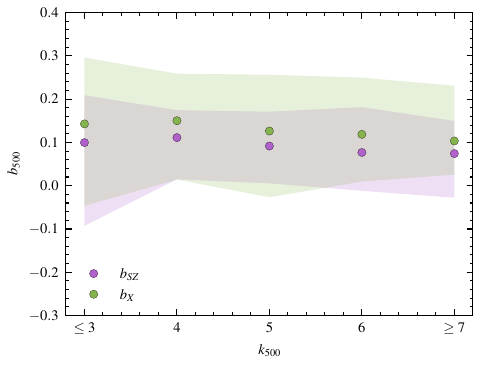}
    \caption{Connectivity and hydrostatic mass bias relation at redshift $z=0$. In green are shown the mass bias median values, $b_{X}$, estimated at $R_{500}$, in violet the mass bias median values, $b_{SZ}$, at $R_{500}$. The shaded areas represent the 16th and 84th percentiles. }
    \label{Fig:k-bias-500}%
\end{figure}

We plot a in Fig. \ref{Fig:k-bias-500} the median values of the hydrostatic mass biases $b_{X}$, in green, and $b_{SZ}$, in violet, computed at $R_{500}$ as a function of the connectivity $k_{500}$. A correlation between the connectivity and the hydrostatic mass bias is not observed. 

\end{appendix}

\end{document}